\documentstyle[12pt]{article}
\def\zid{1\kern-0.36em\llap~1}

\newcommand{\beq}{\begin{equation}}
\newcommand{\ber}{\begin{eqnarray}}
\newcommand{\eeq}{\end{equation}}
\newcommand{\eer}{\end{eqnarray}}

\begin{document}

\begin{titlepage}
\rightline{[SUNY BING 7/1/03 v4] } \rightline{ hep-ph/0308089}
\vspace{2mm}
\begin{center}
{\bf \hspace{0.1 cm} COHERENT PRODUCTION OF PAIRS OF PARABOSONS OF
ORDER 2}\\ \vspace{2mm} Nicholas Frascino and Charles A.
Nelson\footnote{Electronic address: cnelson @ binghamton.edu  }
\\ {\it Department of Physics, State University of New York at
Binghamton\\ Binghamton, N.Y. 13902}\\[2mm]
\end{center}


\begin{abstract}
A parameter-free statistical model is used to study multiplicity
signatures for coherent production of charged-pairs of parabosons
of order $p=2$ in comparison with those arising in the case of
ordinary bosons, $p=1$. Two non-negative real parameters arise
because ``$ab$" and ``$ba$" are fundamentally distinct pair
operators of charge `$+1$', A-quanta and charge `$-1$', B-quanta
parabosons. In 3D plots of ${P_{m}}(q)$$\equiv$ `` the probability
of $m$ paraboson charged-pairs + $q$ positive parabosons" versus
$<n>$ and $<n^2>$, the $p=1$ curve is found to lie on the
relatively narrow 2D $p=2$ surface.
\end{abstract}

\end{titlepage}

\section{Introduction:}

This paper is a theoretical study of multiplicity signatures in
coherent production of charged-pairs of parabosons [1-3] of order
$p=2$ in comparison with those arising in the case of ordinary
bosons, $p=1$. The investigation is partially motivated by
phenomenological coherence analyses, circa 1970, of inelastic $\pi
^{+}\pi ^{-}$ pair production from fixed targets with laboratory
kinetic-energies up to 27 GeV by C.P. Wang [4], and by Horn and
Silver[5]. The present analysis is possible because the
conserved-charge boson coherent states of [5, 6] are analogous to
ones recently constructed for order $p=2$ parabosons [7].  One
physical consequence of ``order $p=2$" is that two or less such
bosons can occupy a totally anti-symmetric state. Appendix 1
contains a brief review of $p=2$ paraboson statistics.

Both the $p=2$ and $p=1$ models considered in this paper are
notably ``statistical'' and crudely ``unrealistic'' in being free
of kinematic and dynamical parameters associated with physically
important production quantities such as the distributions of
available energy/momentum, the kinematic size and other
characteristics of the production region, resonances, masses and
other conserved quantum numbers (e.g. isotopic spin in the $p=1$
case [8]), because only the $U(1)$-charge conservation constraint
has been imposed. Consequently, unlike what some readers would
expect the situation to be from consideration of other
particle-production models, in the figures the relative sizes of
different aspects of the peaks, and of other structures, of the
surfaces/curves cannot be adjusted without additional assumptions
and complications. So despite the authors efforts, an interested
viewer must sometimes force oneself to ``look'' in examination of
some aspects of the figures. In the $p=1$ case, more realistic
models/arguments were made [8] to investigate such issues as
independent-emission and coherence in $\pi ^{+}\pi ^{-}$ pair
production but these analyses are by now, of course, extremely
primitive versus the current very sophisticated computational
treatments [9, 10] of multi-particle production in high-energy
collisions. For instance, even in electron-positron collisions a
wide variety of coherence phenomena such as stage-one and
stage-two spin-correlations in fermion-antifermion pair
production, Bose-Einstein interference effects, and
color-reconnection effects are used in various computational
frameworks to describe and investigate particle production at LEP.
Despite the lack of inclusion of other physics that would be
required in a realistic application, we think that the focus in
this paper on a comparison of these parameter-free statistical
models is interesting and instructive with respect to $p=2$ versus
$p=1$ parastatistics, and to the associated coherences involved in
the production of conserved-charge paraboson pairs.

From the then available data, in [4] Wang reported a regularity
when, for various laboratory kinetic energies of the incoming
primary particle,  the relative frequency of events $P_{m}(q)$ was
plotted against the mean multiplicty $<n>$ for charged $\pi
^{+}\pi ^{-}$ pairs for $pp$, $\pi ^{\pm }p$, $pn$, $\pi ^{\pm
}n,$ and $nn$ collisions.  The data set was a compilation of about
50 inelastic production experiments with incoming primary kinetic
energies extending from below production threshold to $27GeV$.
Wang tried to fit the regularity with two Poisson-type
distributions. In [5], Horn and Silver constructed instead a
simple parameter-free statistical model and found that, ignoring
fluctuations in the data due to the specific structure of the
various channels, the model did agree [5, 7] with the universal
trends of the experimental regularity reported by Wang. One
construction of the Horn-Silver model is/was from the $p=1$
conserved-charge coherent states.  Versus the inelastic reactions'
multiplicity data, in [5] Horn and Silver argued for a statistical
treatment of the gross features because  (i) momentum conservation
should be a weak constraint since the emitted pions occupy a small
part of the available phase space, (ii) total isospin conservation
on the distribution of charged pions should also be weak since
neutral pions are summed over, and so (iii) charge conservation
remains as the important constraint. They derived the $P_{m}(q)$
distribution discussed below.  Their $P_{m}(0)$ distribution, i.e.
the percentage of events with ``m" $\pi ^{+}\pi ^{-}$ pairs and no
extra $\pi^{\pm}$'s , had been considered earlier by Kastrup [11].

At the present time, the only way known to construct the analogous
model in the $p=2$ case is to use conserved-charge coherent
states.  An alternative formulation with respect to either the
statistical aspects or the coherence aspects is not currently
available. So we now briefly digress to discuss what are these
conserved-charge coherent states and then compare them with their
$p=1$ counterpart:

A charged-paraboson pair in order $p=2$ consists of one A quanta
of charge `+1' and one B quanta of charge `-1'.  The Hermitian
charge operator is defined by
\begin{equation}
Q=N_{a}-N_{b}
\end{equation}
where $N_{a,b}$ are the parabose number operators, see Appendix 1.
Although $Q$ does not commute with $a$ or $\ b$, and although the
paraboson pair operators $ab\neq ba$, since
\begin{equation}
\lbrack Q,ab]=0,\quad \lbrack Q,ba]=0,\quad \lbrack ab,ba]=0
\end{equation}
the $p=2$ coherent state can be defined as simultaneous eigenstate of $Q,ab,$ and $%
ba$:
\begin{equation}
Q|q,z,z^{\prime }>=q|q,z,z^{\prime }>,\quad ab|q,z,z^{\prime
}>=z|q,z,z^{\prime }>,\quad ba|q,z,z^{\prime }>=z^{\prime
}|q,z,z^{\prime }>
\end{equation}
Note that unlike in the $p=1$ case (see below), here two complex numbers $z$ and  $%
z^{\prime }$ arise because $ab$ and $ba$ are fundamentally
distinct operators.  Consequently, in the following multiplicity
considerations, two non-negative parameters occur which are the
moduli of these two complex numbers, $u\equiv |z|$ and $v\equiv
|z^{\prime }|$. The explicit expressions for $|q,z,z^{\prime }>$
which involve the modified Bessel functions $I_{[\frac{q}{2}]}(u)$
and $I_{[\frac{q+1}{2}]}(v)$ are given in [7] for $q\geq 0$ and
$q<0$;  note $[\frac{q}{2}]=$ ``integer part of $\frac{q}{2}$''.
For $q \ge 0$, the $M$th-moment of the mean multiplicity for a
$p=2$ charged-paraboson\ pair is $<n^{M}>\equiv <q,z,z^{\prime
}|(N_{b})^M |q,z,z^{\prime }>$.  For $q$ fixed, the percentage of
events with $m$ such pairs, $P_{m}(q)$, is the square of the
moduli of the expansion coefficients of $|q,z,z^{\prime }>$ in
terms of the two-mode parabose number Fock states ( see [7] ). The
explicit multiplicity formulas corresponding to the figures shown
in this paper can be found below in Section 2.

The simpler $p=1$ conserved-charge coherent state $|q,\xi>$
satisfies $Q|q,\xi >=q|q,\xi >$ and $ ab|q,\xi >=\xi |q,\xi
>$. Since $ab$ = $ba$ for $p=1$, $|q,\xi >$ involves a single
complex number $\xi $, and so the Horn-Silver multiplicity
analysis involved a single, non-negative real parameter $x\equiv
|\xi |$.

Because the $p=2$ construction involves two non-negative real
parameters $u$ and $v$, we consider a 3D plot of $P_{m}(q)
=Z(u,v)$ versus the quarter plane defined by $ <n>\equiv <n_{b}> =
X(u,v)$ and $<n^{2}>\equiv <(n_{b})^{2}> = Y(u,v)$.  This is the
obvious generalization of Wang's plot and it is the focus of
Section 2 and its figures. Section 3 contains some additional
multiplicity formulas which are similar in structure for $p=1,2$.
Section 4 contains a brief discussion of some of what has/hasn't
been learned from this analysis of coherent production of
charged-pairs of parabosons of order $p=2$. Appendix 2 and its
figures are focused on how different shaped $\{ u,v \}$ charts, or
coordinate patches, appear when mapped onto the two-dimensional
surface.

\section{ ${P_{m}}(q)$ versus $<n>$, $<n^2>$:}

The analysis in this section is straight-forward because there are
simple explicit mathematical formulas involving modified Bessel
functions and because ``Mathematica" [12] notebooks are convenient
for displaying the parametric 1-parameter $p=1$ ``curve" and
2-parameter  $p=2 $ ``varying-width, folded ribbon." \,We find
that a concise ``curve/ribbon" terminology is useful and
appropriate for describing, for example, Figs. 1a,b, which are for
$m=2$, and $q=1$.  On a 2D side of the 3D display, the $p=2 $
ribbon consequently appears projected as a relatively narrow,
two-dimensional region which will be denoted as a ``band"; see
Fig. 1b which shows the front XZ side of Fig. 1a. In spite of
their being from the more complex $ u \leftrightarrow v$
asymmetric case, these two lead figures have been chosen in order
to {\it ab initio} discuss some details and caveats.

If the reader initially finds Figs. 1a,b and their description too
complex, the reader might go directly to Figs. 2 and 3a,b which
are for the simpler $u \leftrightarrow v$ symmetric case.

In Figs. 1a,b, and in the remaining figures in this paper, the
solid line is the $p=1$ curve.  In these two figures, this $p=1$
curve is also the $\{0,v\}$ line on the $p=2 $ ribbon. Near the
peak, the ``fold" is the bottom edge of the ribbon.  As shown, the
$p=2 $ ribbon consists of open-circles for the non-folded $u \ge
v$ region, and of solid-circles for the folded $ u \le v$ region.
The upper edge of the ribbon is the $\{u,0\}$ set of points.
Slightly to the right of the peak, one can see from the
solid-circles that each line of dots traveling leftward down the
page, bends under (or ``over", whichever as the viewer prefers)
the fold to reach the $\{0,v\}$ line on the ribbon.

For $q$ odd, we assume it is a ``complete fold".  The words
``complete fold" mean that one part of the $ \{ u, v \}$ ribbon
lies exactly on top of another part, and so there is no difference
in going ``under" or ``over" such a fold.

Due to the very limited analytic results, mental/visual care is
generally advisable: (i) Only for the cases of $q=0$ and for $q=1$
is it proven that the $p=1$ curve always lies on the $p=2$ two
dimensional surface. In these two cases the proof appears
un-instructive for then the parametric-solution $u=0, v=2x$ ( and
for $q=0$ also $ u=2x, v=0$ ) avoids confronting different ratios
of modified Bessel functions. For an arbitrary $q$ value, from the
asymptotic limits for the modified Bessel functions, it follows
that as $x \rightarrow 0$, the curve and as $u,v \rightarrow 0$,
the ribbon must begin at the same point: the origin for $m \ne 0$
and $P_{0}(q)=1$ for $m=0$. Similarly, for large parameters for
arbitrary $q$, the curve and ribbon will also be at the same 3D
point with $P_{m}(q) \rightarrow 0$, if $x= \frac {1}{2} (u+v)$.
(ii) Only for $q$ even is it known that there is a complete fold
of the ribbon because then ${P_{m}}(q)$, $<n>$, and $<n^2>$ are
each $u \leftrightarrow v$ symmetric.  For $q$ even, the fold is
the line $u=v$, c.f. the simpler Figs. 2 and 3a,b.

The ``line of dots traveling leftward down the page" discussed
above for the peak region in Figs. 1a,b are a set of $\{ u, v \}$
values from a unit-negative-slope diagonal in the $\{ u, v \}$
domain, see last figure of this paper and its discussion Fig.
8a,b.

Second, the reader should be aware that the ``dots" displayed to
show the two-dimensional surface correspond to specific $\{ u, v
\}$ parameter values and that the associated $u_1, u_2, ...$,
$v_1, v_2, ...$ values are equally spaced, c.f. Fig. 5a below
where some $\{ u, v \}$ coordinate values are shown.  In some
figures, $u_2 - u_1$ is not equal to $v_2-v_1$.  We find that a
careful usage of a few un-connected ``dots" does not mislead in
displaying these two-dimensional surfaces.  The ``dots" should not
be confused with a random generation of data points, such as in a
scatter plot.  Note that the $u$ and $v$ parameters do not map
into an orthogonal coordinate chart on the ribbon.  In general, at
each point on the ribbon the curvature is non-zero.

    Third, in consideration of Figs. 1a,b, as well as the
other figures in this paper, it should be noted that, even near
the peak, the width of the ribbon is narrow versus, for instance,
the half-width of the $p=1$ peak.  This fact, besides the folding
and the role of the $u$ and $v$ parameters, would be an important
issue in attempting empirically to distinguish coherent production
pairs of $p=2$ parabosons versus those of ordinary $p=1$ bosons.
On the other hand, the ribbon width is indeed non-zero and to the
eye there is generally a systematic above, or below, displacement
of the visual center of the $p=2$ ``band" from the $p=1$ curve as
in Figs. 1a,b.

For the $m=2$, and $q=1$ case shown in Figs. 1a,b the formulas for
the $p=1$ curve are
\begin{eqnarray}
<n>^{(1)}=\frac{xI_{2}(2x)}{I_{1}(2x)},\quad <n^{2}>^{(1)}=<n>^{(1)}+\frac{%
x^{2}I_{3}(2x)}{I_{1}(2x)} \\ P_{2}^{(1)}(1)
=\frac{x^{5}}{12I_{1}(2x)}
\end{eqnarray}
and for the $p=2$ ribbon are
\begin{eqnarray}
<n>^{(2)}=\frac{1}{2}(\frac{uI_{1}(u)}{I_{0}(u)}+\frac{vI_{2}(v)}{I_{1}(v)}%
) \\
<n^{2}>^{(2)}=<n>^{(2)}+\frac{1}{4}(\frac{u^{2}I_{2}(u)}{I_{0}(u)}+\frac{%
2uvI_{1}(u)I_{2}(v)}{I_{0}(u)I_{1}(v)}+\frac{v^{2}I_{3}(v)}{I_{1}(v)})
\\ P_{2}^{(2)}(1)
=\frac{3u^{4}v+6u^{2}v^{3}+v^{5}}{384I_{0}(u)I_{1}(v)}
\end{eqnarray}
As in (4-8), a superscript ``$(1)$ or $(2)$'', for $p=1$ or $p=2$,
can be respectively written on $<n>$, $<n^{2}>$, and $P_{m}(q)$
when needed to avoid confusion. Usually one knows the $p$ order
from the working context, and so these superscripts can often be
suppressed.

Before discussing other specific $m$ and $q$ cases, we consider
the parametric formulas for an arbitrary curve and ribbon.

\subsection{$p=1$ curve: }

The $p=1$ curve $(X(x),Y(x),Z(x))$ is parametrized by the real
non-negative parameter $x$ .  For $q$ $\geq 0$, in terms of
modified Bessel functions, $I_{q}(2x)$,
\begin{eqnarray}
X(x) &\equiv &<n>=\frac{xI_{q+1}(2x)}{I_{q}(2x)} \\ Y(x) &\equiv
&<n^{2}>=<n>+\frac{x^{2}I_{q+2}(2x)}{I_{q}(2x)}
\end{eqnarray}
Equivalently, $<n^{2}>=x^{2}(1-\frac{qI_{q+1}(2x)}{xI_{q}(2x)})$
follows due to the recursion relation for $I_{q}$'s. The
probability for `` $m$ boson pairs + $q$ positive bosons '' is
\begin{equation}
Z(x)=P_{m}(q)=\frac{x^{2m+q}}{I_{q}(2x)m!(m+q)!}
\end{equation}
For example, in the analysis of ``ideal'' data for the production
of purely multi-pion final states, $P_{m}(q)$ would be the
probability for the production of $``m+q"$ $\pi ^{+}$'s and $``q"$
$\pi ^{-}$'s. For $p=1$, it is instructive to consider the
statistically-fundamental
\begin{equation}
P_{m}(q)=(N_{q})^{2}\frac{x^{2m}}{m!(m+q)!}
\end{equation}
with an $m$-independent normalization constant
$(N_{q})^{-2}=x^{-q}I_{q}(2x)$ because in this way one sees that
it is via normalization that the modified Bessel function appears
in development of the simple idea of
charge-conservation-constrained Poisson distributions for
independent $\pi^+$ and $\pi^-$ production [5, 8].

For multi-pion final states Wang [1], and later Horn and Silver
[2] in consideration of (12) as a model, plotted $P_{m}(q)$ versus
$<n>$ for small values of $m$ and $q$. \ This is the $XZ$ plane,
or ``front'' projection of the 3D figures considered in this
paper.

\subsection{$p=2$ surface: }

The $p=2$ two-dimensional surface $(X(u,v),Y(u,v),Z(u,v))$ is
parametrized by the two real non-negative parameters $u$ and $v$.
$\ $As for the $p=1$ curve discussed above, only the
$Z$-coordinate is $m$ dependent.

($q$ Even): \ The case of $q$ even and $q$ $\geq 0$ is symmetric in $u$ $%
\leftrightarrow $ $v$ with $Z(u,v)=Z(v,u)$, see below, and the $u$ $%
\leftrightarrow $ $v$ symmetric
\begin{eqnarray}
X(u,v) &\equiv &<n>^{(2)}=\frac{1}{2}\left( \frac{uI_{\frac{q+2}{2}}(u)}{I_{%
\frac{q}{2}}(u)}+\frac{vI_{\frac{q+2}{2}}(v)}{I_{\frac{q}{2}}(v)}\right)
\\
Y(u,v) &\equiv &<n^{2}>^{(2)}=<n>^{(2)}+\frac{1}{4}\left( \frac{u^{2}I_{%
\frac{q+4}{2}}(u)}{I_{\frac{q}{2}}(u)}+2uv\frac{I_{\frac{q+2}{2}}(u)}{I_{%
\frac{q}{2}}(u)}\frac{I_{\frac{q+2}{2}}(v)}{I_{\frac{q}{2}}(v)}+\frac{%
v^{2}I_{\frac{q+4}{2}}(v)}{I_{\frac{q}{2}}(v)}\right)
\end{eqnarray}

($q$ Odd): \ The case of $q$ odd and $q$ $\geq 0$ is asymmetric in $%
u$ $\leftrightarrow $ $v$ with $Z(u,v)\neq Z(v,u)$, and the $u$ $%
\leftrightarrow $ $v$ asymmetric
\begin{eqnarray}
X(u,v) &\equiv &<n>^{(2)}=\frac{1}{2}\left( \frac{uI_{\frac{q+1}{2}}(u)}{I_{%
\frac{q-1}{2}}(u)}+\frac{vI_{\frac{q+3}{2}}(v)}{I_{\frac{q+1}{2}}(v)}\right)
\\
Y(u,v) &\equiv &<n^{2}>^{(2)}=<n>^{(2)}+\frac{1}{4}\left( \frac{u^{2}I_{%
\frac{q+3}{2}}(u)}{I_{\frac{q-1}{2}}(u)}+2uv\frac{I_{\frac{q+1}{2}}(u)}{I_{%
\frac{q-1}{2}}(u)}\frac{I_{\frac{q+3}{2}}(v)}{I_{\frac{q+1}{2}}(v)}+\frac{%
v^{2}I_{\frac{q+5}{2}}(v)}{I_{\frac{q+1}{2}}(v)}\right)
\end{eqnarray}

For $q \ge 0$, the probability for `` $m$ paraboson pairs + $q$
positive parabosons '' is
\begin{eqnarray}
Z(u,v) &\equiv &P_{m}^{(2)}(q)=(N_{q}^{(2)})^{2}\sum_{i=1}^{m+1}\tilde{P}%
_{q,m;i} \\
(N_{q}^{(2)})^{-2} &=&(\frac{u}{2})^{-[\frac{q}{2}]}I_{[\frac{q}{2}]}(u)(%
\frac{v}{2})^{-[\frac{q+1}{2}]}I_{[\frac{q+1}{2}]}(v) \\
\tilde{P}_{q,m;i} &=&\frac{u^{2r}v^{2s}}{2^{2m}[\frac{m+i}{2}]![\frac{q+m+i}{%
2}]![\frac{m+1-i}{2}]![\frac{q+m+1-i}{2}]!}
\end{eqnarray}
where in $\tilde{P}_{q,m;i}$
\begin{eqnarray}
r &\equiv &[\frac{m-(-)^{q+m+i}i}{2}+\frac{1-(-)^{q}}{4}] \\ s
&\equiv &[\frac{m+(-)^{q+m+i}i}{2}+\frac{1+(-)^{q}}{4}]
\end{eqnarray}
In paraboson statistics, it is sometimes convenient to use $[x]=$
``integer part of $x$'', and $[x]!=[x][x-1]\cdots 1$, $[0]!=1$.
Note that this $[x]$ symbol occurs in (18-21).  The use of the
symbol $[x]$ enables a compactification of several analytic
expressions and its use is convenient in computational notebooks.
However,
it is sometimes instructive, as in (13-16), to write out the various cases by letting $q=2Q$%
, or $2Q+1$; $Q=0,1,\cdots $; or similarly for $m$ or $i$.
Throughout this paper, the appearance of ``square braces'' always
denotes such an integer truncation. Note that the summation over
``i" in (17) is also due to the fact that $ab \neq ba$ [7].  The
compact expressions are
\begin{eqnarray*}
<n>^{(2)}=\frac{1}{2}\left( \frac{uI_{[\frac{q+2}{2}]}(u)}{I_{[\frac{q}{2}%
]}(u)}+\frac{vI_{[\frac{q+3}{2}]}(v)}{I_{[\frac{q+1}{2}]}(v)}\right)
\\
<n^{2}>^{(2)}=<n>^{(2)}+\frac{1}{4}\left( \frac{u^{2}I_{[\frac{q+4}{2}]}(u)%
}{I_{[\frac{q}{2}]}(u)}+2uv\frac{I_{[\frac{q+2}{2}]}(u)}{I_{[\frac{q}{2}]}(u)%
}\frac{I_{[\frac{q+3}{2}]}(v)}{I_{[\frac{q+1}{2}]}(v)}+\frac{v^{2}I_{[\frac{%
q+5}{2}]}(v)}{I_{[\frac{q+1}{2}]}(v)}\right)
\end{eqnarray*}

Since from (17) the $p=2$ case non-trivially involves two
parameters,
$u$ and $v$, the figures in this paper are 3D ones with $%
P_{m}^{(p)}(q)\equiv Z^{(p)}$ plotted versus the two-dimensional
$<n>^{(p)}$ $\equiv X^{(p)}$ and $<n^{2}>^{(p)}$ $\equiv Y^{(p)}$
plane.  Only the first octant is used.

\subsection{Other $m$ and $q$ cases: }

For the case $m=0$, $q=0$ the parametric expressions are very
simple:
\begin{eqnarray}
<n>^{(1)}=\frac{xI_{1}(2x)}{I_{0}(2x)},\quad <n^{2}>^{(1)}=<n>^{(1)}+\frac{%
x^{2}I_{2}(2x)}{I_{0}(2x)} \\ P_{0}^{(1)}(0) =\frac{1}{I_{0}(2x)}
\end{eqnarray}
\begin{eqnarray}
<n>^{(2)}=\frac{1}{2}(\frac{uI_{1}(u)}{I_{0}(u)}+\frac{vI_{1}(v)}{I_{0}(v)}%
)\quad  \\
<n^{2}>^{(2)}=<n>^{(2)}+\frac{1}{4}(\frac{u^{2}I_{2}(u)}{I_{0}(u)}+\frac{%
2uvI_{1}(u)I_{1}(v)}{I_{0}(u)I_{0}(v)}+\frac{v^{2}I_{2}(v)}{I_{0}(v)})
\\ P_{0}^{(2)}(0) &=&\frac{1}{I_{0}(u)I_{0}(v)}
\end{eqnarray}
This is the simplest $u \leftrightarrow v$ symmetric case and the
``right" $YZ$-side is shown in Fig. 2.  As previously mentioned,
in the symmetric case, the fold is along $u=v$ which is shown by
the dashed line in Fig. 2.  The two edges of the original ribbon,
$\{0,v\}$ and $\{u,0\}$ are respectively mapped by $v=2x$, $u=2x$,
into the $p=1$ curve. Only open-circle points from the $u \ge v$
region are shown.  All $m=0$ curves/ribbons drop down from $P_{o}
(q)=1$.

Figs. 3a,b, respectively show the 3D plot and the right $YZ$-side
for another symmetric case $m=1$, $q=0$.  Note that, unlike in the
preceding Fig. 2, in this case the $p=1$ curve now lies on the top
of the ribbon until there is an almost $180^o$ twist, after the
peak where the $p=1$ curve is on bottom.  In Fig. 3b the $YZ$-side
quite clearly shows this twist. The parametric expressions are
respectively $ P_{1}^{(1)}(0) =\frac{x^{2}}{I_{0}(2x)}$ , $
P_{1}^{(2)}(0) =\frac{u^{2}+v^{2}}{4I_{0}(u)I_{0}(v)}$.

In appendix 2, there is a discussion of the $\{u,v\}$ coordinate
charts associated with the asymmetric case $m=1$, $q=1$ shown in
Figs. 4a - 4d.  The parametric equations are respectively $
P_{1}^{(1)}(1) = \frac{x^{3}}{2I_{1}(2x)}$, $ P_{1}^{(2)}(1) =
\frac{u^{2}v+v^{3}}{8I_{0}(u)I_{1}(v)}$.  [ The omitted figures
for the asymmetric case $m=0$, $q=1$ similarly show the fold
occurring in the $u \le v$ region. ] Fig. 4a is the 3D display and
Fig. 4b is the projection onto the front $XZ$ plane. Fig. 4c is
from a different 3D viewpoint.  It provides a view back into the
origin, with a $YX$ floor below and a $YX$ ceiling above, in a
``left-to-right, front axes labeling". This figure shows how
nearly vertical the ribbon is. Lastly, Fig. 4d shows a close-up of
the $m=1$, $q=1$ peak. This figure illustrates the fold.  In it
the upper solid line is the $p=1$ curve and the lower solid line
is the $u=v$ curve which separates the open-circles ($u \ge v$)
from the solid-circles ($u \le v$).

The different shaped $\{ u,v \}$ charts of Appendix 2 can be used
in investigating the folding which occurs in a transverse crossing
of the ribbon.  Similarly, the lines $u=0$, also $v=0$ for
$q$-odd, $u=v$, and for $q \ge 2$ also the independent $p=1$
curve, can be used.  Both techniques are useful in studying what
occurs in going out along the ribbon. In proceeding from the
origin, or for $m=0$ from $P_0 (q)=1$, there are twist(s) and a
peak (absent for $m=0$). If as in Fig. 4c, the ribbon is viewed
back in from the high $u$ and $v$ parametric coordinates, then in
moving out from the origin but viewing back towards the origin,
the almost $180^o$ twisting of the $ u \ge v $ edge from the peak
onward is counter-clockwise for $(m,q)= (1,0),(1,1),(2,1)$. For $
(2,1), (2,2), (2,3)$ there is also a twist near the origin. For
the peak in $(1,1)$, c.f. Fig. 4a, the $Z$ coordinate values are:
for the $v=0$ top edge, $Z(2.65,0)=0.475$; for the $p=1$ curve,
$Z(0,3.25)=0.436$; for the $u=v$ curve $Z(1.85,1.85)=0.42$; and
for the fold $Z(1.5, 2.25) = 0.41$.

\section{Additional Multiplicity Formulas:}

In the $p=2$ case, the $M$-th moment of the mean multiplicity
\begin{equation}
<n_{b,a}^{M}>=(N_{q})^{2}(\widehat{D_{b,a}})^{M}(N_{q})^{-2}
\end{equation}
where $\widehat{D_{b}}=\frac{1}{2}(u\frac{\partial }{\partial u}+v\frac{%
\partial }{\partial v}),\widehat{D_{a}}=\widehat{D_{b}}+q$. \ It follows that
\begin{eqnarray}
<n>=<n_{b}>=(N_{q})^{2}\mathit{S}_{1}(u,v), \\
<n^{2}>=<n_{b}^{2}>=(N_{q})^{2}(\mathit{%
S}_{2}+\mathit{S}_{1}), \\
<n_{b}^{3}>=(N_{q})^{2}(\mathit{S}_{3}+3\mathit{S}_{2}+\mathit{S}_{1}),
\\
<n_{b}^{4}>=(N_{q})^{2}(\mathit{S}_{4}+6\mathit{S}_{3}+7\mathit{S}_{2}+%
\mathit{S}_{1}), \\
<n_{b}^{5}>=(N_{q})^{2}(\mathit{S}_{5}+10\mathit{S}_{4}+25\mathit{S}_{3}+15%
\mathit{S}_{2}+\mathit{S}_{1}), ...
\end{eqnarray}
where, with ${ M \choose l}$ a binomial coefficient,
\begin{eqnarray}
\mathit{S}_{M}(u,v) &=&\ \bar{c}\sum_{l=0}^{M} { M \choose l} (\frac{u}{2}%
)^{M-l}(\frac{v}{2})^{l}I_{[\frac{q+2(M-l)}{2}]}(u)I_{[\frac{q+1+2l}{2}]}(v)
\\
\bar{c} &=&2^{[\frac{q}{2}]+[\frac{q+1}{2}]}u^{-[\frac{q}{2}]}v^{-[\frac{q+1%
}{2}]}
\end{eqnarray}
The function $\mathit{S}_{M+1}(u,v)$ is generated by $(\widehat{D_{b}}-M)%
\mathit{S}_{M}(u,v)=\mathit{S}_{M+1}(u,v)$.  The expression for
$(N_{q}^{(2)})^{2}$ is (18) above.

Note that as in the $p=1$ case,
\begin{equation}
<n_{a}^{M}>=\sum_{t=0}^{M} { M \choose t} q^{t}<n_{b}^{M-t}>
\end{equation}
When $q=0$, $<n_{a}^{M}>=<n_{b}^{M}>$, and $\ $for fixed $q$,
$<n_{a}>=<n_{b}>+q$.

For ($q=-|q|)<0$, there are the following relations to the $q>0$
expressions:
\begin{equation}
<n_{a}^{M}>_{-|q|}=<n_{b}^{M}>_{|q|}
\end{equation}
in (27-32) and exchange $ u\leftrightarrow v $,
\begin{equation}
<n_{b}^{M}>_{-|q|}=<n_{a}^{M}>_{|q|}
\end{equation}
in (27, 35) and exchange $ u\leftrightarrow v$; and
$P_{m}(|q|)=P_{m+|q|}(-|q|)$.

The analogous formulas for the $p=1$ case are
\begin{equation}
<n_{b,a}^{M}>^{(1)}=(N_{q}^{(1)})^{2}(\widehat{D_{b,a}^{(1)}}%
)^{M}(N_{q}^{(1)})^{-2}
\end{equation}
where $\widehat{D_{b}^{(1)}}=\frac{1}{2}(x\frac{\partial
}{\partial x}), \widehat{D_{a}^{(1)}}=\widehat{D_{b}^{(1)}}+q$. So
$ <n>^{(1)}=<n_{b}>^{(1)}=(N_{q}^{(1)})^{2}{s}_{1}(x)$ ,
\newline $<n_{b}^{2}>^{(1)}=(N_{q}^{(1)})^{2}({s}_{2}+{s}_{1}), ... $
where
\begin{equation}
\mathit{s}_{M}(x)=\ x^{M-q}I_{q+M}(x)
\end{equation}
The function ${s}_{M+1}(x)$ is generated by
$(\widehat{D_{b}^{(1)}}-M) {s}_{M}(x)={s}_{M+1}(x)$.  The
expression for $(N_{q}^{(1)})^{2}$ is given after (12) above.

For ($q=-|q|)<0$ for $p=1$, $<n_{a}^{M}>_{-|q|}=<n_{b}^{M}>_{|q|}$
, $ <n_{b}^{M}>_{-|q|}=<n_{a}^{M}>_{|q|}$, and
$P_{m}(|q|)=P_{m+|q|}(-|q|)$.

\section{Discussion:}

The 2D surface occurs in the 3D plots of the relative
probabilities $P_{m}(q) $ versus $<n>$, $<n^{2}>$ as required by
the coherences embodied in the equations in (3) for the $p=2$
conserved-charge coherent states. \ \ Unlike the analogous $p=1$
case, where such a model was motivated from a regularity in $\pi
^{+}\pi ^{-}$ data, it is necessary in the case of the production
of the charged-pairs of  parabosons of order $p=2$ to assume an
analogous coherent production mechanism and to assume a
reasonable, but definite, treatment of the $ab$ and $ba$
operators. \ \ Neither of these two assumptions might be true if
paraboson pairs are found to be produced in nature in high energy
physics collisions, e.g. in central-diffactive-exchange
experiments [14], or in some area of contemporary
astrophysics/cosmology, e.g. in dark matter detectors [15] or by
very high energy cosmic rays [16]. \ In the infra-red domains of
QED and of QCD, coherent-state/degenerate-state coherence is a
well-known phenomena. \ On the other hand in QCD, partonic jets
dominate the very high energy hadronic production processes. \ In
the case of the lower energy $\pi ^{+}\pi ^{-}$ production, as was
emphasized by Horn-Silver, and others, the situation is complex
and there is much more physics and phenomenological structure than
that incorporated in the reference $p=1$ model considered in this
paper.  In the approximation in which one neglects such additional
physics, this $p=1$ model is parameter free because the $x$
parameter is effectively replaced by $<n>$, the mean number of
final charged pairs. Similarly, the analogous $p=2$ model is
parameter free for the $u$ and $v$ parameters are effectively
replaced by $<n>$ and $<n^2>$. At the present time, the existence
and relative importance of other conserved quantum numbers, of
resonances or other phenomenological interaction effects, etc. is
completely unknown for a coherent production of pairs of
parabosons of order $2.$ Nevertheless, with respect to experiments
[14-16], one conclusion from this paper is that from a
parameter-free statistical model, one would expect a signature of
relatively narrow bands, due to projection of varying-width folded
ribbons, will be present in 2D analysis of pair multiplicities
from coherent production of parabosons of order $2$.

In the case of $p=2$ parabosons, there will be a second kinematic
variable [4] and multiplicity data will not scale in terms of a
single-variable curve [11]. For instance, if the variable for the
$p=1$ curve is $x \sim \frac{1}{2}(u+v)$ , then for $p=2$
parabosons there will also be a dependence on $y \sim
\frac{1}{2}(u-v)$. In Fig. 8a, with $x \sim \frac{1}{2}(u+v)$
fixed, the sensitivity to the second variable $y \sim
\frac{1}{2}(u-v)$ is very striking and significantly greater than
the naive width of the ribbon.  In the region of the peak, Figs.
1b and 4b show that this signature is considerably enhanced when
there is an extra positive paraboson, versus the case of only
charged-paraboson-pairs, Fig. 3b. While an explicit dynamical
production model is required to theoretically investigate the
sensitivity with respect to specific kinematic variables, in
independent-particle-emission models the total energy in the
emitted particles is monotonically related to the intensity
strength of the source. This suggests that two simple and useful
kinematic variables are the sum, $E_{total} = {E^{+}}_{total} +
{E^{-}}_{total} $, and difference ${E^{+}}_{total} -
{E^{-}}_{total} $, of the total emitted $Q= 1$ and $Q=-1$
energies.

The ``identical particle'' defining parabose tri-linear relations
leave the two pair operators, $ab$ and $ba$,  fundamentally
distinct which (3) and present analysis maintain by the use of the
two distinct $u$ and $v$ parameters. In the $p=1$ case, the
$x^{2}=x_{+}x_{-}$ parameter can be interpreted as a product of
the intensity strengths of the $\pi ^{+}$ and $\pi ^{-}$ sources.
In the $p=2$ case, $u$ and $v$ can be interpreted as the intensity
strengths of the ``$ab$" and ``$ba$" sources. Their difference is
due to the two distinct orderings of the $a$ and $b$ operators.
The physical observables are not always $u$ $\leftrightarrow $ $v$
symmetric.  In particular, the presence of a $q$ odd total charge
produces an asymmetry in the folding of the 2D surface; for
instance, see Eqs. (15,16) and (8).

There is overall $A^{+} \leftrightarrow B^{-} $ symmetry ( $U(1)$
charge symmetry ): The ``Probability for $(m + |q|) $ $A^{+}$'s
and $(m)$ $B^{-}$'s " equals ``Probability for $(m) $ $A^{+}$'s
and $(m +
|q| )$ $B^{-}$'s ", because $P_{m}(|q|)=P_{m+|q|}(-|q|)$ and $%
<n_{b}^{M}>_{-|q|}=<n_{a}^{M}>_{|q|}$ for $M=$integer.

The $U(1)$ charge might be a ``hidden conserved-charge'' such as
to yield pure pair production, i.e. only final state events with
$q=0$ would occur such as in the production of strange particles
via the strong interactions. \ In this case, or by a designed
selection of only pair final state events, the $p=2$ \ versus
$p=1$ results of this paper can still be used with
$\sum_{m=0}^{\infty }P_{m}(0)=1$. \ This is the simpler symmetric
case and because of $q=0$\ there\ would be more analytic control
in such an analysis.

{\bf Acknowledgments: }

One of us (CAN) thanks experimental and theoretical physicists for
discussions.  We thank Ted Brewster, Lorene Evans, Anthony Poole,
and James L. Wolf for computer support. This work was partially
supported by U.S. Dept. of Energy Contract No. DE-FG 02-86ER40291.

\newpage

{\bf Appendix 1:  Parabosons of order 2 }

In local, relativistic quantum field theory, identical particles
obey either (i)  parabose and parafermi statistics [1-3] for which
the number of particles in an antisymmetric or a symmetric state,
respectively, cannot exceed a given integer $p$, or (ii) for two
space dimensions, infinite statistics based on the braid group
[13]. In parabose statistics, instead of bilinear, there are
fundamental trilinear commutation relations:
\begin{equation}
\begin{array}{c}
\lbrack a_k,\{a_l^{\dagger },a_m\}]=2\delta _{kl}a_m, \;
[a_k,\{a_l^{\dagger },a_m^{\dagger }\}]=2\delta _{kl}a_m^{\dagger
}+2\delta _{km}a_l^{\dagger },
\\
\lbrack a_k,\{a_l,a_m\}]=0, \hspace{2pc} (k,l,m=1,2)
\end{array}
\end{equation}
where $[C,D]\equiv  CD-DC$, and $\{C,D\} \equiv CD+DC$. In the
case of only two kinds of parabosons, there are some simple
commutation relations between ``A'' paraboson operators and the
``B'' paraboson operators: letting $a\equiv a_1,$ $b\equiv a_2$,
\begin{equation}
[a,b^2]=0, \; [b,a^2]=0, \; [a^{\dagger },b^2]=0, \; [b^{\dagger
},a^2]=0
\end{equation}
plus the hermitian conjugate relations.  Order $p=2$ is simpler
because there is the ``self-contained set" of 3 relations [2]
\begin{eqnarray}
a_{m}a_{l}a_{k}^{\dagger }-a_{k}^{\dagger }a_{l}a_{m} &=&2\;\delta
_{kl}\;a_{m} \\ a_{k}a_{l}^{\dagger }a_{m}-a_{m}a_{l}^{\dagger
}a_{k} &=&2\;\delta _{kl}\;a_{m}-2\;\delta _{lm}\;a_{k} \\
a_{k}a_{l}a_{m}-a_{m}a_{l}a_{k} &=&0
\end{eqnarray}
The parabose number operators for $p=2$ order are defined by
\begin{equation}
N_a=\frac 12\{a^{\dagger },a\}-1, \; N_b=\frac 12\{b^{\dagger
},b\}-1
\end{equation}
In [7], the state-vector space for two-mode parabosons and the
order $p=2$ conserved-charged parabose coherent states were
constructed.

Production and decay selection rules [2, 3] exclude known
particles from obeying other [1] than the usual boson and fermion
statistics, i.e. order $p=1$ parastatistics.

{\bf Appendix 2:  $\{u,v\}$ Charts on 2D surface }

In a determination of the statistical uncertainties for an
application of the parametric formulas in Section 2 in comparison
with actual experimental data, the structure(s) of the folded,
partially-twisted ribbons versus $p=1$ curves will enter and have
to be correctly treated. For this reason, and for possible more
abstract theoretical/mathematical use by other readers, the
following details about how different shaped $\{u,v\}$ coordinate
patches appear when mapped onto the 2D surface should be of
interest.  The lines $u=0$, also $v=0$ for $q$-odd, and $u=v$
provide useful standard references in examining what is taking
place on each ribbon.  In addition, for $q \ge 2$ the $p=1$ curve
is a different reference since it no longer is the $u=0$ line.

The $m=1$, $q=1$ case which is $u \leftrightarrow v$ asymmetric is
used in this appendix.  This case is both generic and visually
simple. Figs. 4a-4d were earlier displays of this case.   The
discussion begins with the non-folded $ u \ge v $ region:

Figs. 5a,b show how the $u \ge v $ ``strip" from $u=2$ to $u=3$ is
mapped onto the peak region of the ribbon.  In 3D Fig. 5a, the
solid line that lies on the bottom of the ``patch" on the ribbon
is the $u=v$ set of points.  The open-circle points on the top of
the patch are the $v=0$ set of points.  The points on the origin
side in Fig. 5a are for the vertical $u=2$ line in Fig. 5b. The
points on the patch farthest from the origin, which appear to have
a nearly constant negative slope, are for the other vertical line
in Fig. 5b, i.e. for $u=3$.  Here there have been 5 sets of
vertical points mapped from Fig. 5b onto the 5a patch on the
ribbon.

Figs. 6a,b similarly show the differences when the $u \ge v $
points arise from three more widely separated constant-$u$
vertical lines with respectively $ u \simeq 2, 3, 4$.  To the
left-side of the peak, $u \simeq 2$, the spacing between these
constant-$u$-value points is rather small, whereas on the right of
the peak, where $u \simeq 4$, the spacing between such points is
significantly greater.  The points are equally spaced in $v$ so
there are more points taken at $u \simeq 4$ than at $u \simeq 2$.

Horizontal ``strips" can similarly be used to study the folded $u
\le v$ region.  However, a ``ring" region or a
``unit-negative-slope strip" region can be used which includes
both $u \ge v$ and $u \le v$ domains. Figs. 7a,b show what occurs
when points in a ``ring" of radii $r_{<}=3$ and $r_{>}=4$ are
mapped onto the ribbon.  In 7a, the open-circle-dots are again
from the simpler $u \ge v$ region, whereas the solid-dots are from
the $u \le v$ region where the fold occurs.  Fig. 7c shows a
close-up where the $u=0$ points ($p=1$ curve) are on the upper
solid curve; the $u=v$ points which divide the ring in half are on
the lower solid curve. [ If instead, the points are from
pie-slice-like sectors, the associated lines on the ribbon then
extend indefinitely on out the ribbon. ]

Lastly, Fig. 8a,b show what happens when points are taken from a
somewhat similar, but sometimes more useful, unit-negative-slope
strip in the $u,v$ quadrant. In Fig. 8a the solid line is the
$u=v$ curve.  The ``line of dots traveling leftward down the page"
discussed in the text for the peak region in Figs. 1a,b are a set
of $\{ u, v \}$ values from such a unit-negative-slope diagonal.
If instead, the points are taken from the orthogonal
unit-positive-slope strips, the associated lines on the ribbon
will start on a $\{ 0, v \}$ or $\{ u, 0 \}$ line, depending on
which side of $ u=v $ the strip is from, and then the line will
extend indefinitely on out the ribbon.

\begin{center}
{\bf Figure Captions}
\end{center}

Fig. 1a:  The probability ${P_{2}}(1)$ of ``2
charge-paraboson-pairs plus 1 positive paraboson" versus $<n>$,
$<n^2>$. This is the more complex $ u \leftrightarrow v$
asymmetric case. The solid line is the $p=1$ curve. Near the peak,
the fold is the bottom edge of the ribbon.  The open-circles are
for the non-folded $u \ge v$ region, and the solid-circles for the
folded $u \le v$ region, see discussion in Sec. 2.

Fig. 1b: The front projection, or $XZ$-side, of Fig. 1a, so
${P_{2}}(1)$ versus $<n>$.

Fig. 2: The right-side projection, or $YZ$-side, for the
probability ${P_{0}}(0)$  of ``zero charge-paraboson-pairs plus
zero additional single charged parabosons" versus $<n^2>$.  Since
$q$ is even, this is the simpler $ u \leftrightarrow v$ symmetric
case.  Only open-circles from the $u \ge v$ region are shown.  The
fold is the dashed line $u=v$.

Fig. 3a:  The probability ${P_{1}}(0)$ of ``1
charge-paraboson-pair plus zero additional charged parabosons
versus $<n>$, $<n^2>$.  This is also the simpler $ u
\leftrightarrow v$ symmetric case so only open-circles from the $
u \ge v$ region need to be shown.  The solid line is the $p=1$
curve.  The fold is the dashed line $u=v$.

Fig. 3b: The right projection, or $YZ$-side, of Fig. 3a, so
${P_{1}}(0)$ versus $<n^2>$.

Fig. 4a:  The probability ${P_{1}}(1)$ of ``1
charge-paraboson-pairs plus 1 positive paraboson" versus $<n>$,
$<n^2>$.  This case is $ u \leftrightarrow v$ asymmetric.

Fig. 4b: The front projection, or $XZ$-side, of Fig. 4a, so
${P_{1}}(1)$ versus $<n>$.

Fig. 4c:  A 3D view ``back into the origin" of Fig. 4a.

Fig. 4d: A close-up view of the peak of Fig. 4a.  Appendix 2 and
its figures discuss various $\{ u, v \}$ coordinate charts for
this asymmetric case.

Fig. 5a:  For ${P_{1}}(1)$ of Figs. 4, the image of the $u \ge v$
``strip" from $u=2$ to $u=3$ shown in Fig. 5b.

Fig. 5b: The domain $u \ge v$ ``strip" from $u=2$ to $u=3$ for
Fig.5a.

Fig. 6a:  For ${P_{1}}(1)$, the image of the three vertical lines
in Fig. 6b.

Fig. 6b: The three vertical lines $u \simeq 2, 3, 4$ which are
mapped to the 2D surface in Fig. 6a.

Fig. 7a:  For ${P_{1}}(1)$, the image of the ``ring" of points of
Fig. 7b.

Fig. 7b: The domain of points in the ``ring" of radii $r_{<} =3$
and $r_{>} = 4$ which are mapped into Fig. 7a.

Fig. 7c: A close-up of Fig. 7a.

Fig. 8a:  For ${P_{1}}(1)$, the image of the points in the
unit-negative-slope domain in Fig. 8b.  The $\{ u, v \}$ values
shown are a subset from the unit-negative-slope line of points
nearest the origin in Fig. 8b.  Note the location of the end
points $\{ 4, 0 \}$ and $\{ 0, 4 \}$.

Fig. 8b: The unit-negative-slope domain whose points are mapped
into  Fig. 8a:

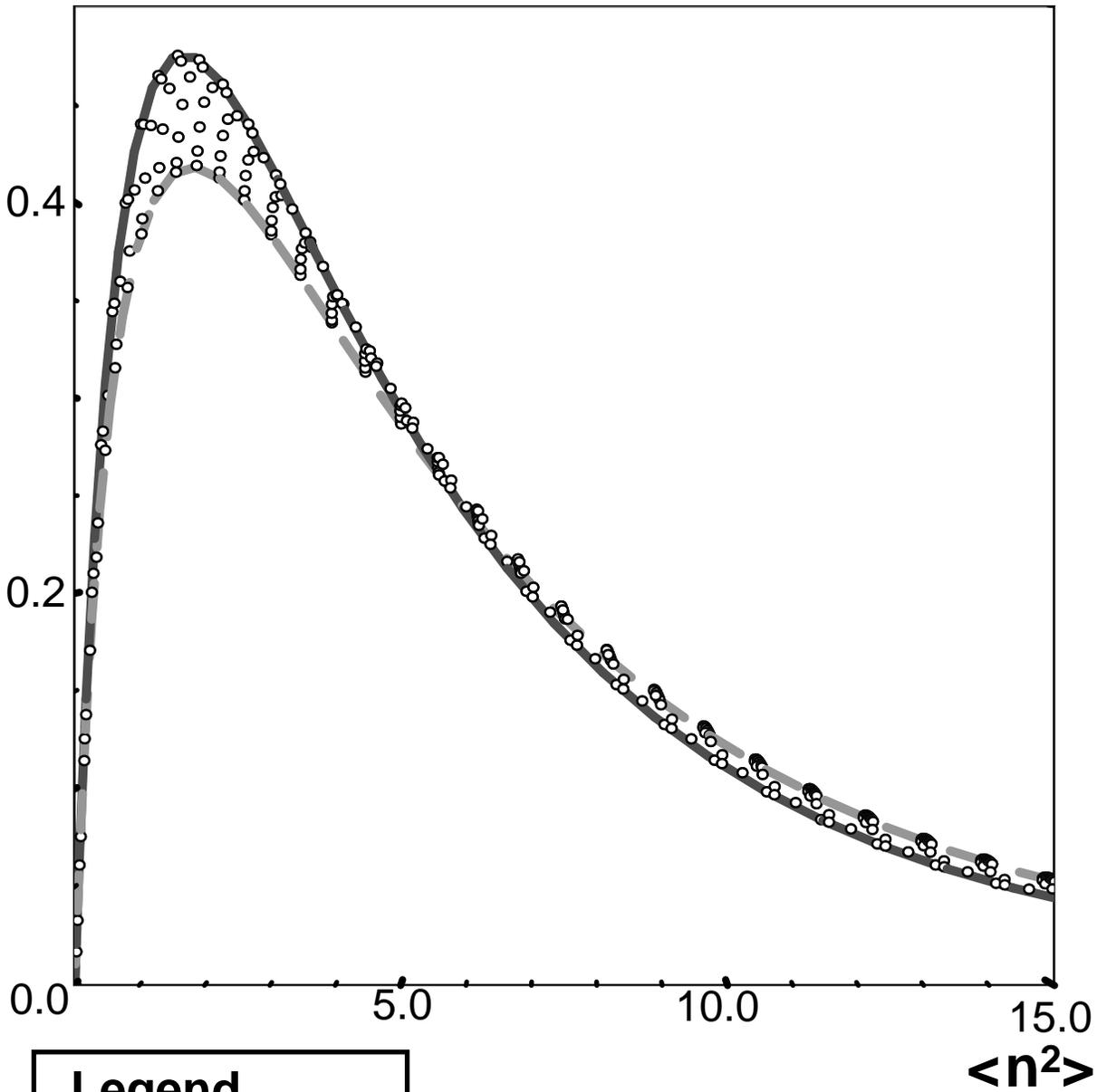
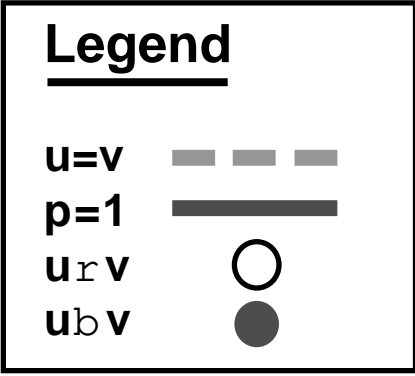
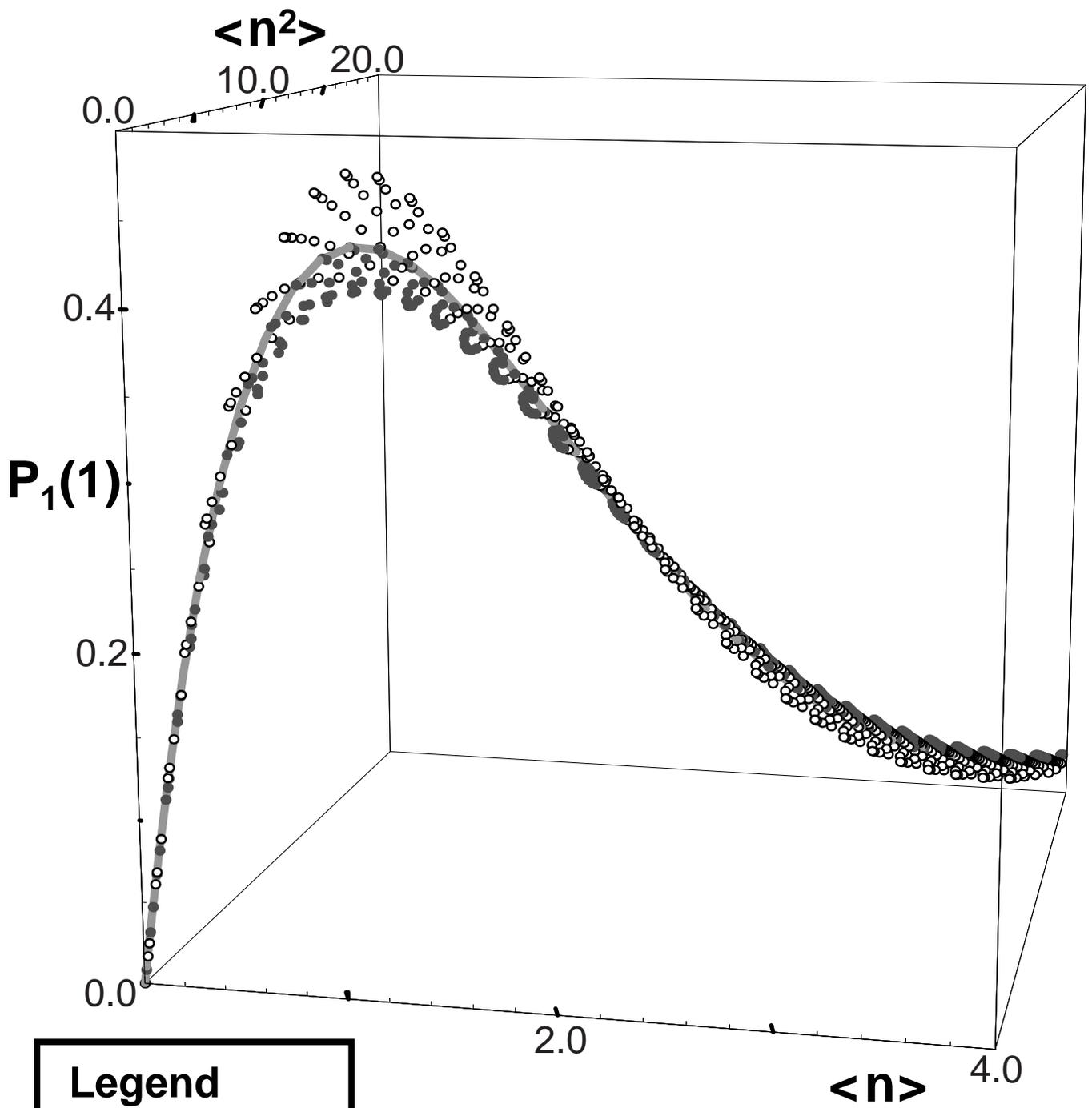
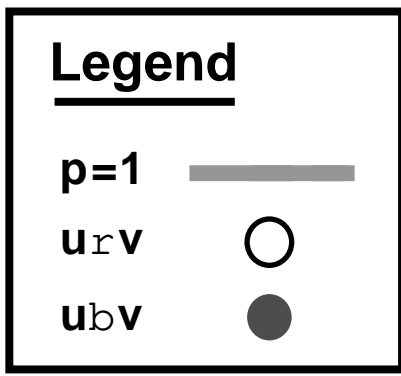
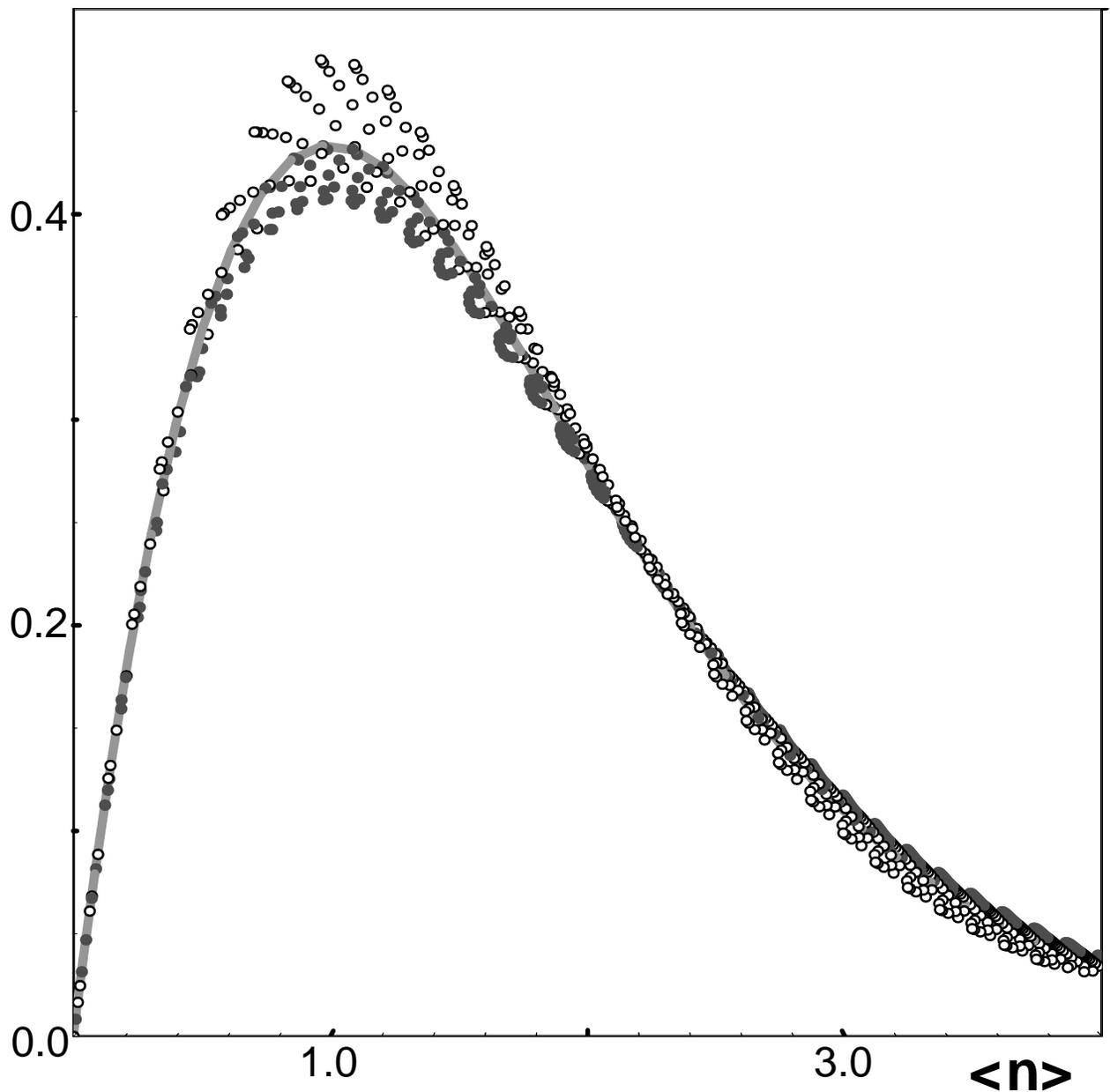
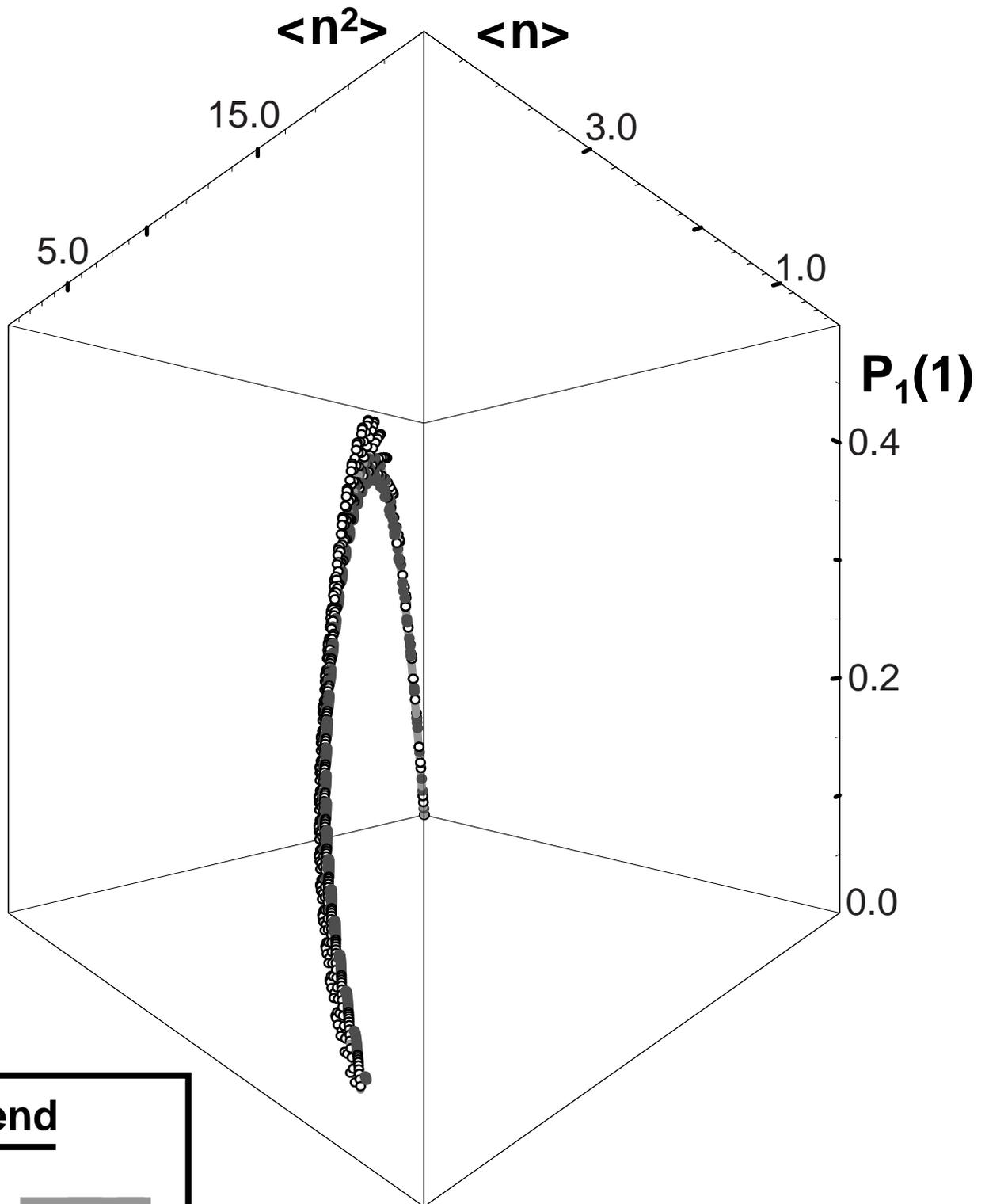
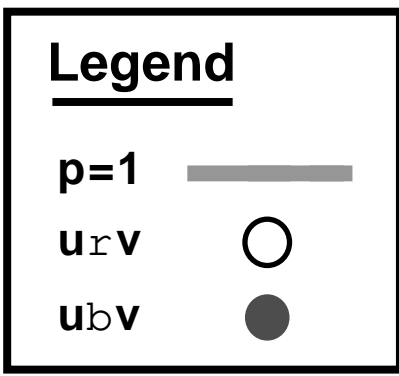
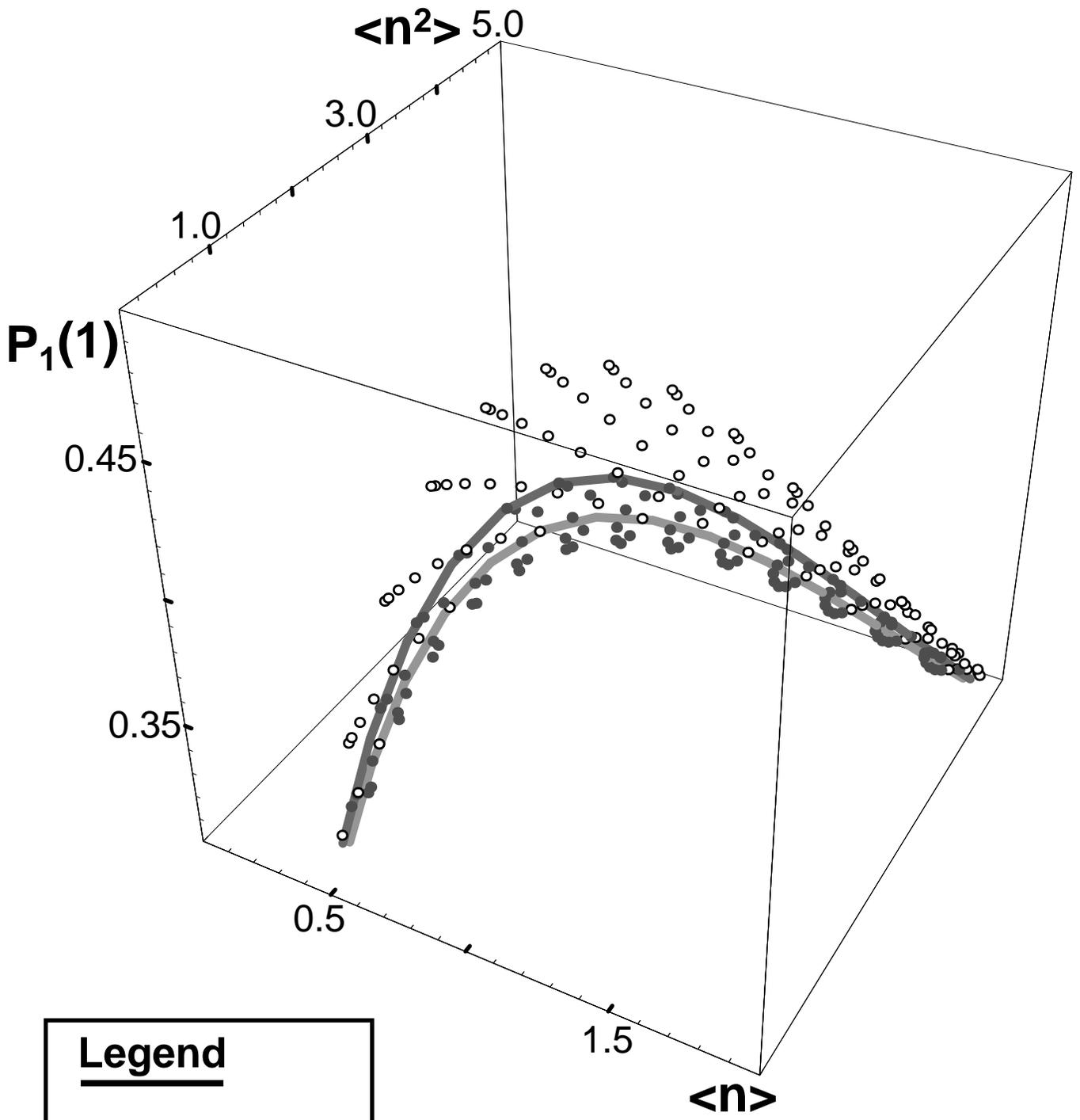
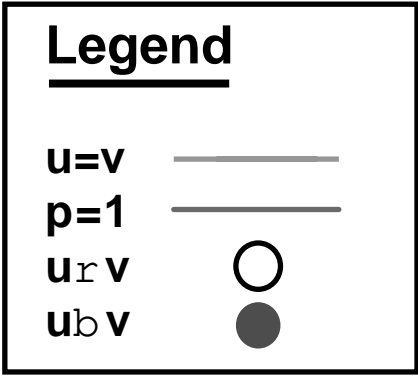
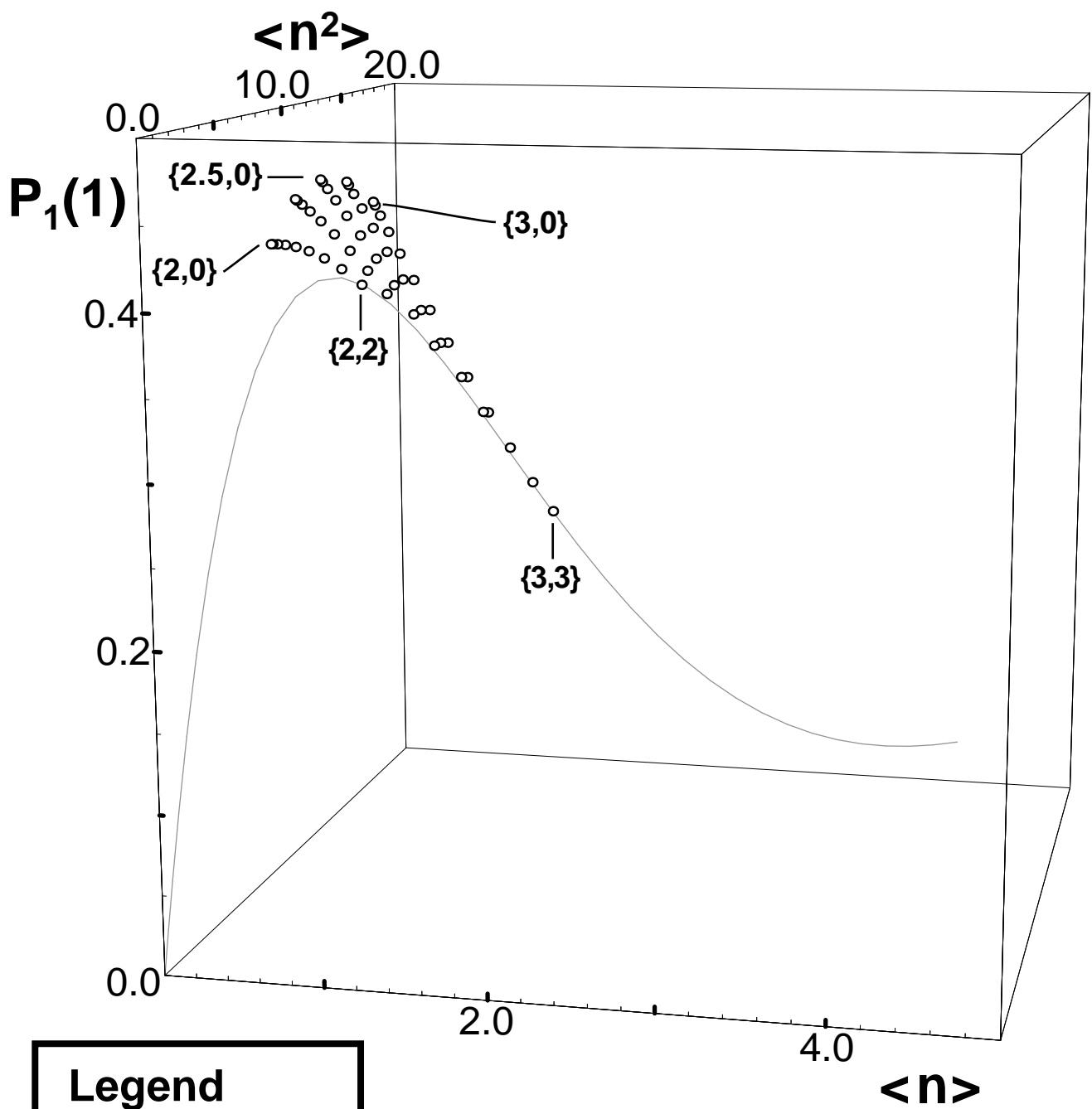
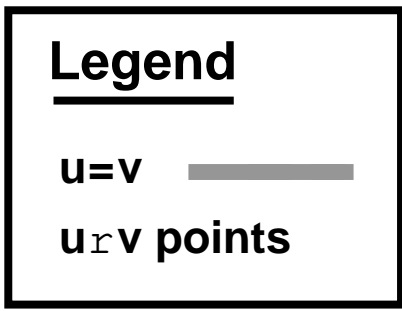
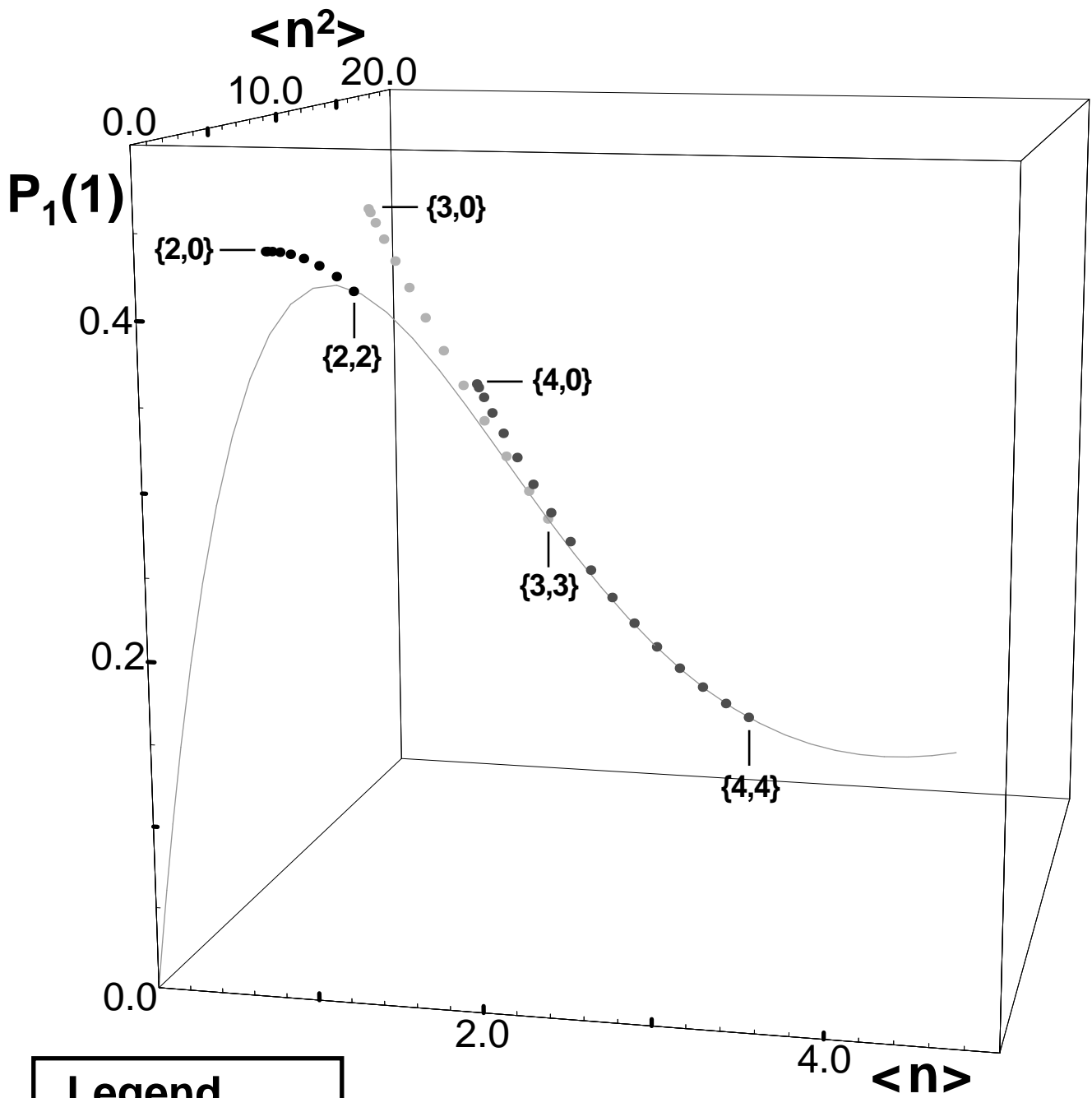
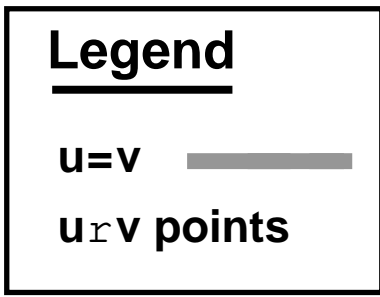
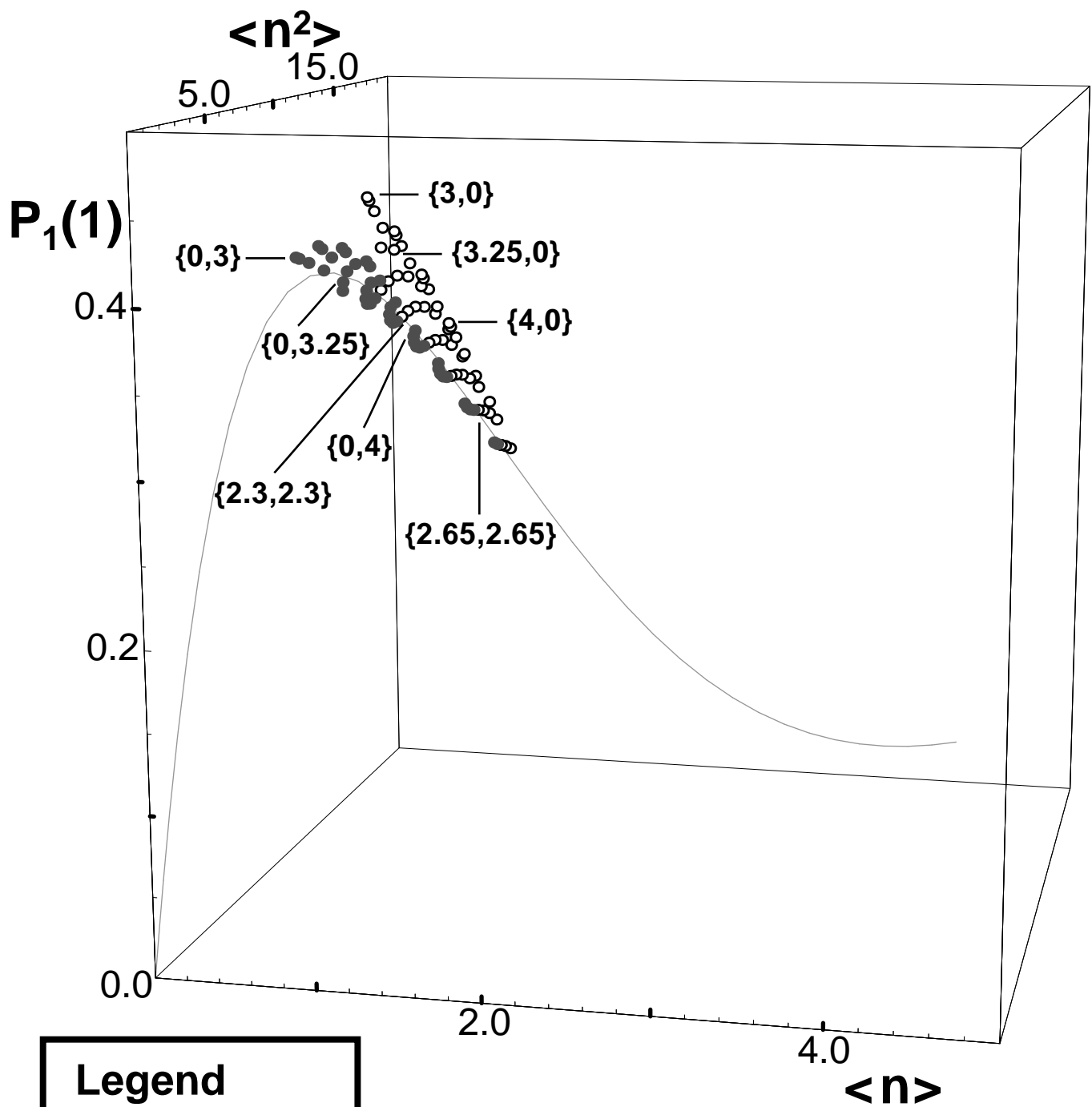
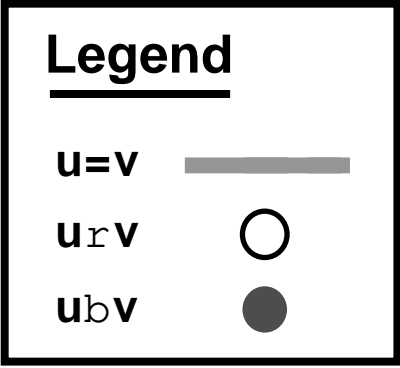
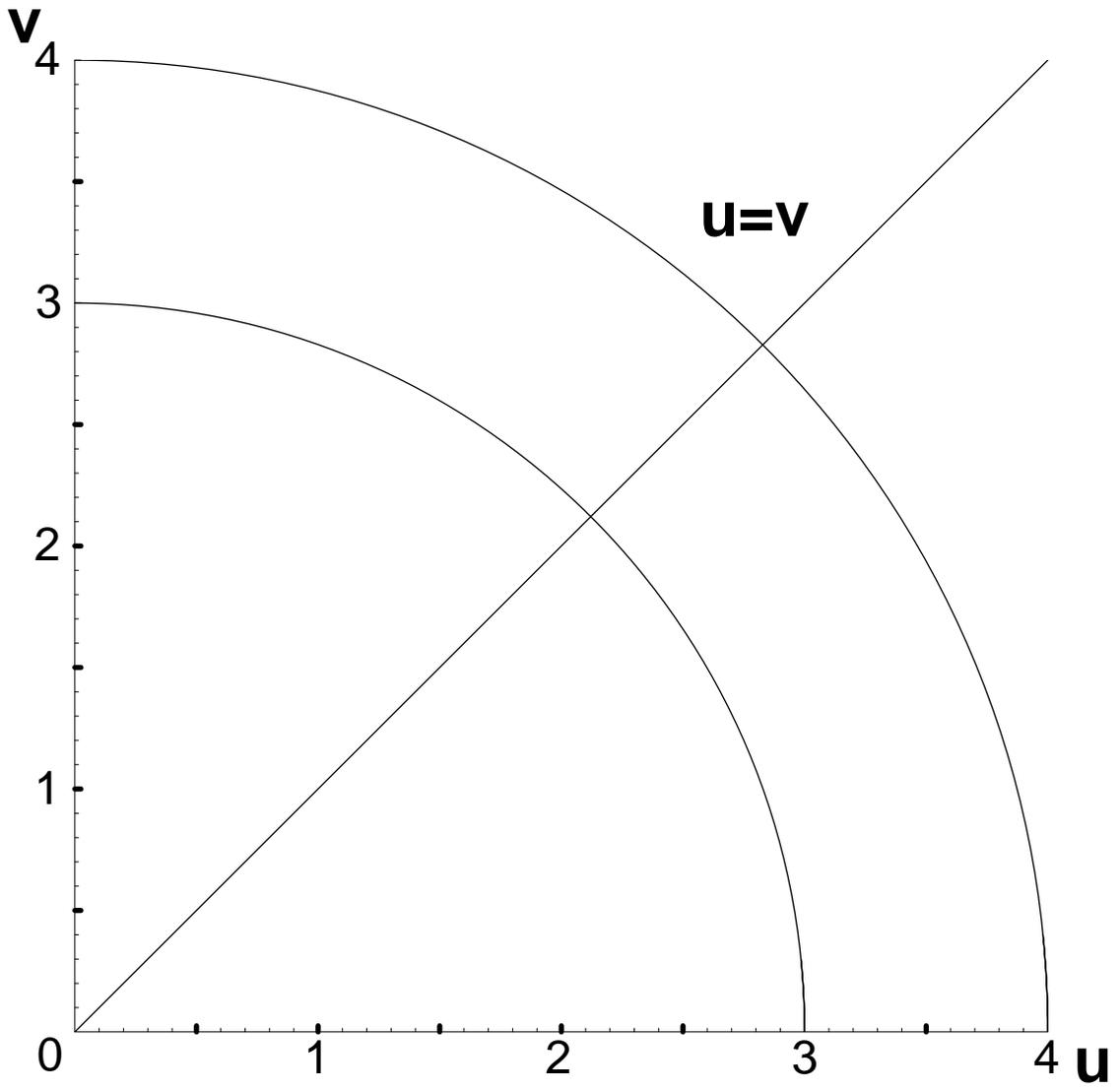
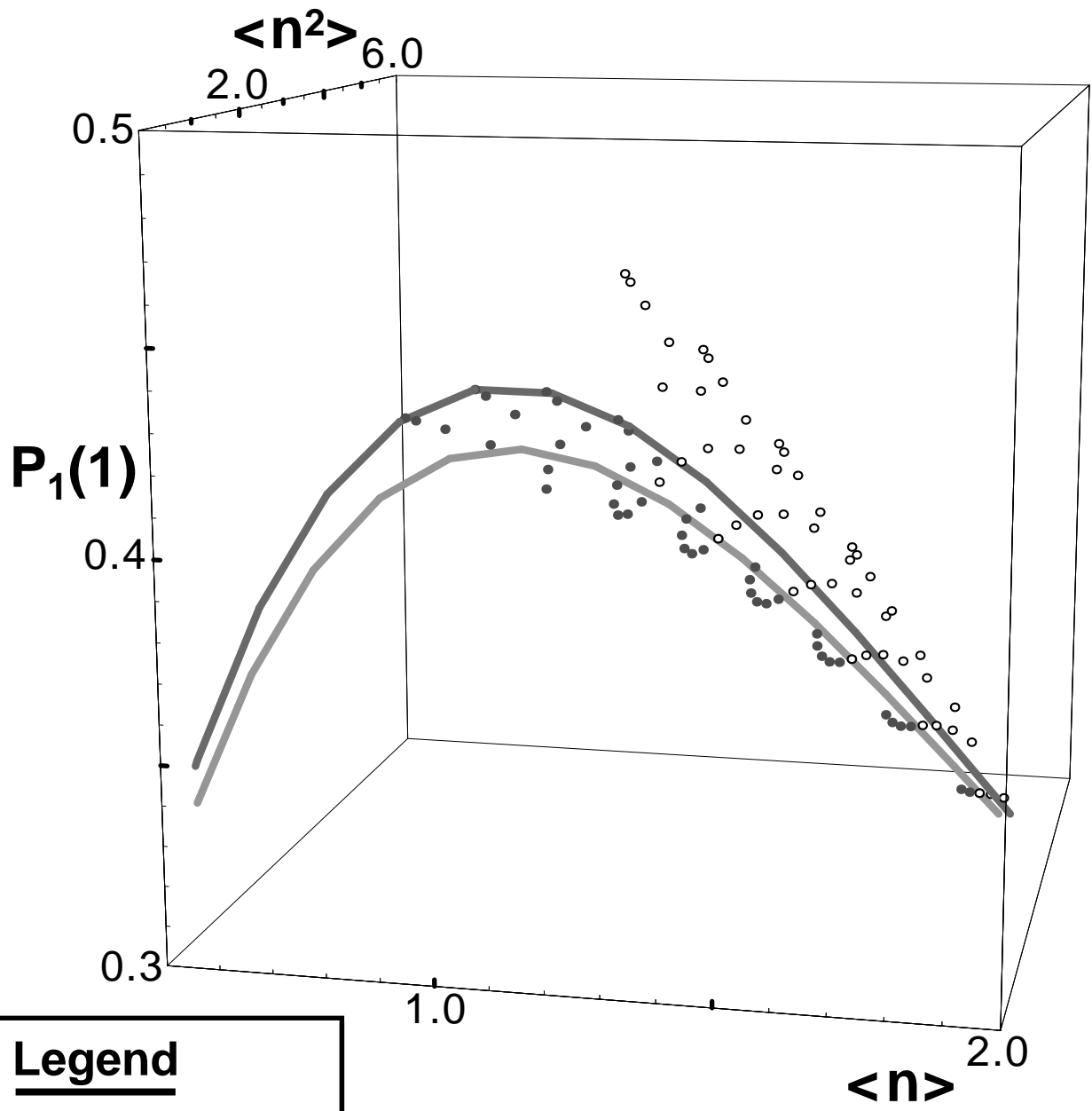
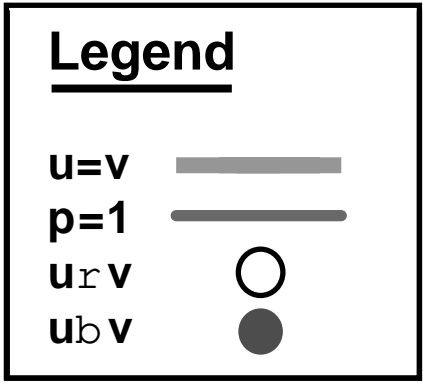
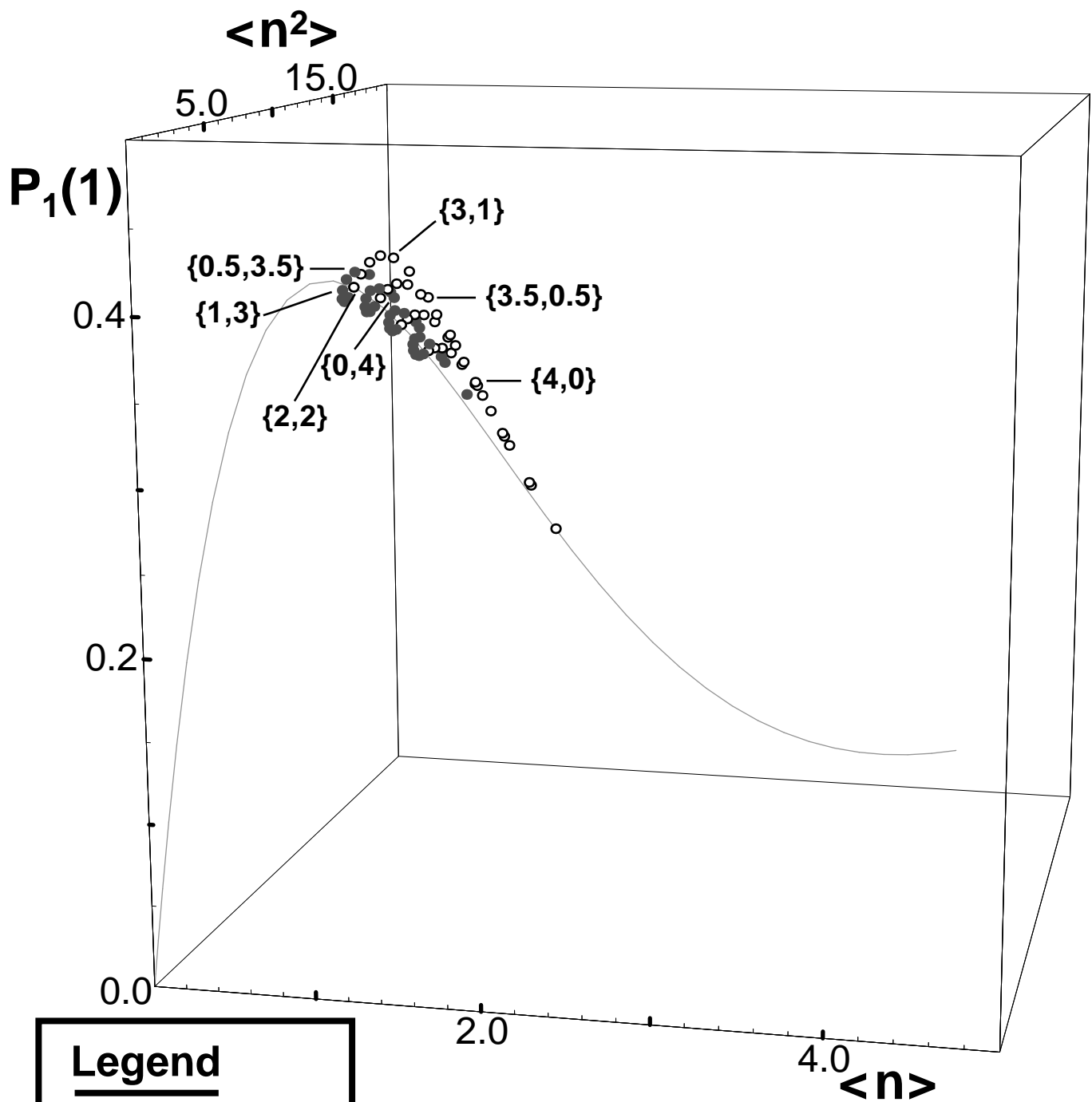
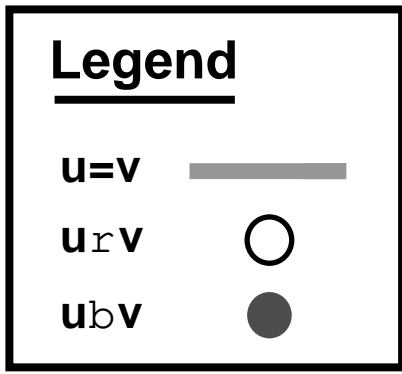
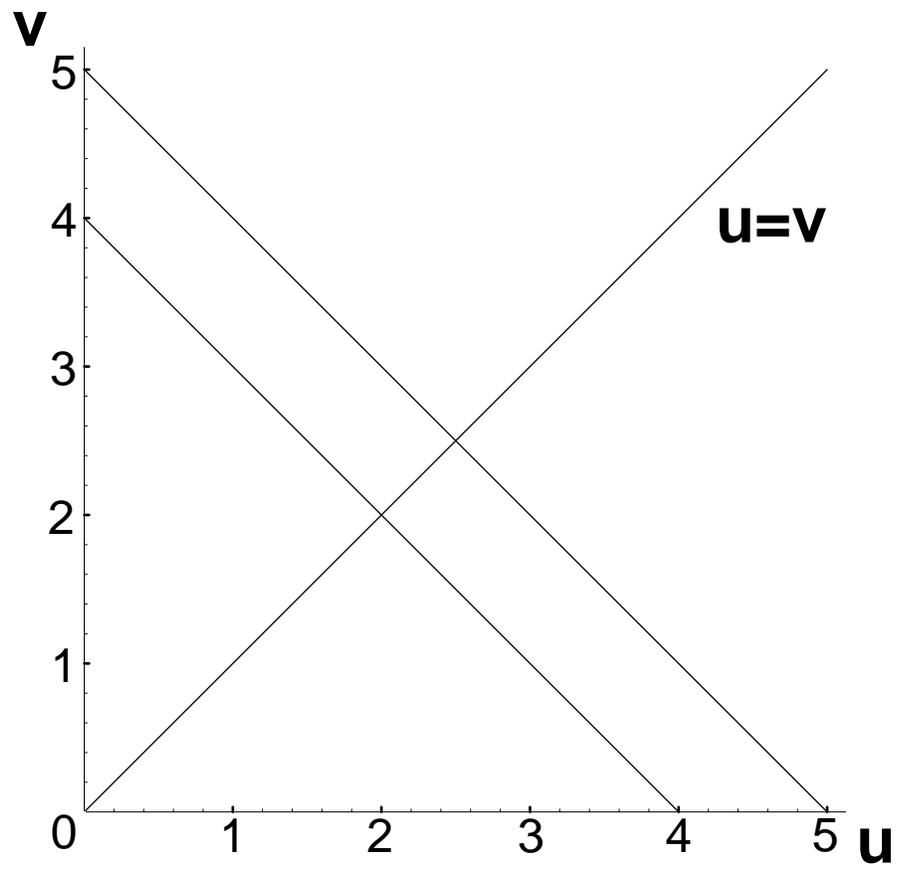
\end{document}